\documentclass[prx,twocolumn,amsmath,amssymb,superscriptaddress,longbibliography]{revtex4-2}

\usepackage[pdftex]{graphics}
\usepackage{graphicx}
\usepackage{times}
\usepackage{amsmath}
\usepackage{xcolor}
\usepackage{color}

\begin{document}
\title{Quasimolecular $J_{\rm tet}$\,=\,3/2 moments in the cluster Mott insulator GaTa$_4$Se$_8$}

\author{M. Magnaterra}
\affiliation{Institute of Physics II, University of Cologne, 50937 Cologne, Germany}
\author{J. Attig}
\author{L. Peterlini}
\affiliation{Institute for Theoretical Physics, University of Cologne, 50937 Cologne, Germany}
\author{M. Hermanns}
\affiliation{Department of Physics, Stockholm University, AlbaNova University Center, SE-106 91 Stockholm, Sweden}
\author{M. H. Upton}
\author{Jungho~Kim}
\affiliation{Advanced Photon Source, Argonne National Laboratory, Argonne, Illinois 60439, USA}
\author{L.~Prodan}
\affiliation{Experimental Physics V, Center for Electronic Correlations and Magnetism, 
	University of Augsburg, Germany}
\author{V.~Tsurkan}
\affiliation{Experimental Physics V, Center for Electronic Correlations and Magnetism, 
	University of Augsburg, Germany}
\affiliation{Institute of Applied Physics, Moldova State University, MD 2028, Chisinau, R. of Moldova}
\author{I.~Kézsmárki}
\affiliation{Experimental Physics V, Center for Electronic Correlations and Magnetism, 
	University of Augsburg, Germany}
\author{P.H.M. van Loosdrecht}
\affiliation{Institute of Physics II, University of Cologne, 50937 Cologne, Germany}
\author{M. Gr\"{u}ninger}
\affiliation{Institute of Physics II, University of Cologne, 50937 Cologne, Germany}

\date{September 21, 2023}

\begin{abstract}
Quasimolecular orbitals in cluster Mott insulators provide a route to tailor exchange 
interactions, which may yield novel quantum phases of matter.  	
We demonstrate the cluster Mott character of the lacunar spinel GaTa$_4$Se$_8$ using 
resonant inelastic x-ray scattering (RIXS) at the Ta $L_3$ edge. 
Electrons are fully delocalized over Ta$_4$ tetrahedra, forming quasimolecular 
$J_{\rm tet}$\,=\,3/2 moments. 
The modulation of the RIXS intensity as function of the transferred momentum $\mathbf{q}$
allows us to determine the cluster wavefunction, which depends on competing intracluster 
hopping terms that mix states with different character. This mixed wavefunction is decisive 
for the macroscopic properties since it affects intercluster hopping and exchange interactions 
and furthermore renormalizes the effective spin-orbit coupling constant.   
The versatile wavefunction, tunable via intracluster hopping, opens a new perspective on the 
large family of lacunar spinels and cluster Mott insulators in general.
\end{abstract}

\maketitle

With strong spin-orbit coupling, novel forms of quantum magnetism may 
emerge from unconventional magnetic moments that exhibit exotic exchange couplings. 
The Kitaev spin liquid is a prominent example \cite{Kitaev06,Hermanns18}.
Bond-directional Kitaev exchange has been realized in, e.g., $5d^5$ honeycomb iridates 
with spin-orbit-entangled $j$\,=\,1/2 moments \cite{Jackeli09,Rau16,Trebst22,Chun15,Magnaterra23}. 
Another intriguing case is given by $5d^1$ $j$\,=\,3/2 moments on an \textit{fcc} lattice, 
e.g., in double perovskites. These moments experience bond-dependent multipolar interactions, 
giving rise to a rich phase diagram that includes multipolar order and a chiral quantum 
spin liquid with Majorana fermion excitations \cite{Chen10,Natori16,Romhanyi17,Tehrani23}.

Exchange-coupled local moments exist in Mott insulators, where electrons are localized 
on individual sites. A new flavor is offered by cluster Mott insulators, which can be 
viewed as the electronic equivalent of a molecular crystal \cite{Pocha2000,AbdElmeguid2004,Sheckelton12,Chen14,Lv15,Khomskii21}.
In these, electrons occupy quasimolecular orbitals that are delocalized over a cluster, 
e.g., a dimer or trimer, while inter\-cluster charge fluctuations are suppressed 
by Coulomb repulsion. 
The emerging quasimolecular magnetic moments are the fundamental units determining 
the macroscopic low-energy properties. 
Importantly, the character of these moments can be tuned by internal degrees of freedom.
%One example is given by the Ir$_2$O$_9$ dimers with three holes in, e.g., 
%Ba$_3$InIr$_2$O$_9$, which has been discussed as a spin-liquid candidate 
%with persistent spin dynamics down to 20\,mK \cite{Dey17}. 
One example is an Ir$_2$O$_9$ dimer with three holes as in the spin-liquid candidate 
Ba$_3$InIr$_2$O$_9$ \cite{Dey17}. With increasing intradimer hopping, 
the dimer moments change from $J_{\rm dim}$\,=\,1/2 to 3/2 \cite{Li20,Revelli_In}. 
In general, the quasimolecular wavefunction depends on 
competing intracluster hopping terms and is highly sensitive to the cluster shape. 
The ability to tune intracluster hopping via external or chemical pressure offers 
the promising perspective to tailor the moments and thereby the character and symmetry 
of intercluster exchange interactions with the aim to realize novel magnetic quantum 
phases of matter.

\begin{figure}[t]
	\centering
	\includegraphics[width=\columnwidth]{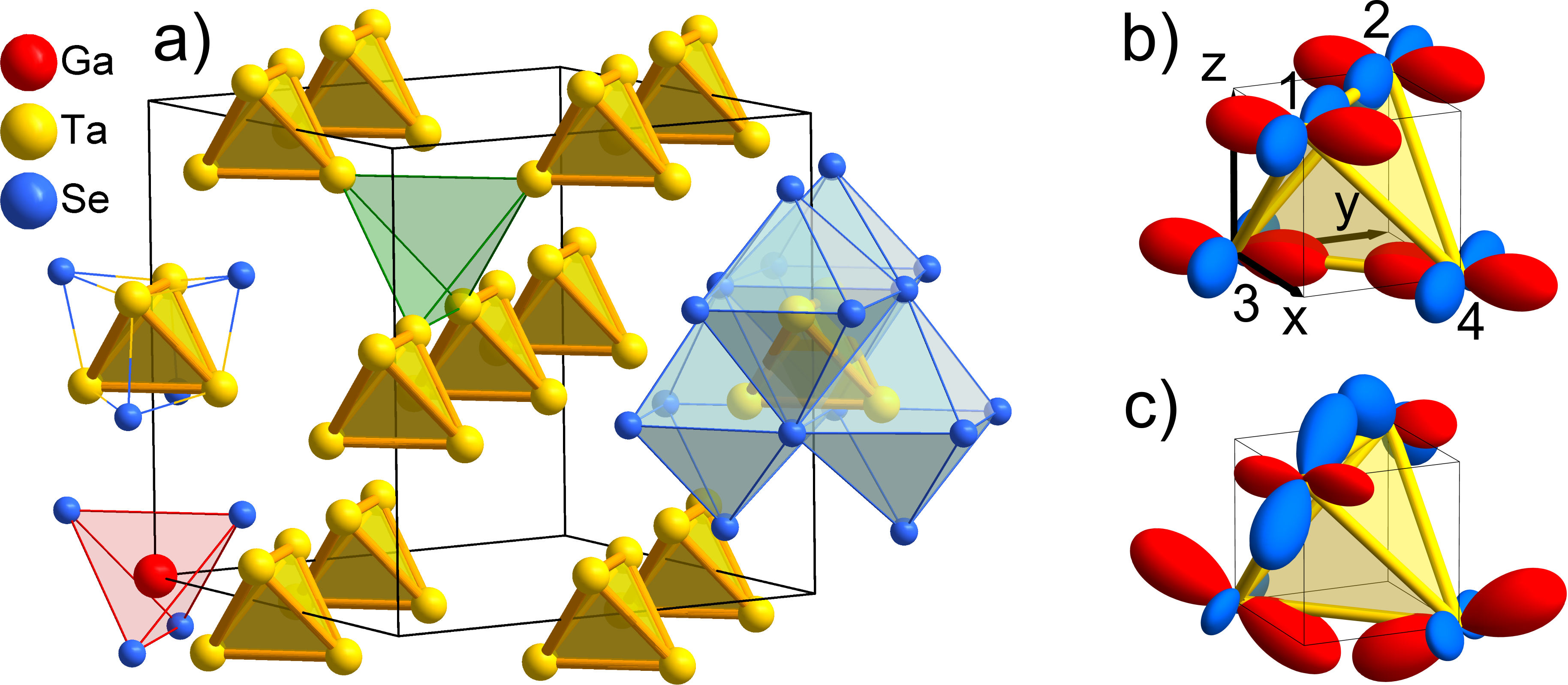}
	\caption{a) Simplified sketch of cubic GaTa$_4$Se$_8$ \cite{Pocha2005}.	
	Not all Ga and Se ions are shown. 
	The structure corresponds to a NaCl-like lattice of tetrahedral (GaSe$_4$)$^{-5}$ 
	(red) and heterocubane (Ta$_4$Se$_4$)$^{+5}$ units. 
	Tetrahedral Ta$_4$ clusters (yellow) arise from edge-sharing TaSe$_6$ octa\-hedra (blue) 
	and form an \textit{fcc} lattice.
	The intracluster Ta-Ta distance $d$\,=\,3.0\,\AA\ is much shorter than the intercluster 
	one ($4.3$\,\AA, edges of green tetrahedron). 
	b) Bonding quasimolecular $xy_b$ orbital, see Eq.\ (\ref{eq:xyB}). 
	c) $t_2(xy)$ orbital with sizable antibonding character, 
	see Eq.\ (\ref{eq:t2AB}) for $\alpha$\,=\,2.
	}
	\label{fig:structure}
\end{figure}

We focus on the transition-metal $M_4$ tetrahedra in the large family of lacunar spinels 
$AM_4X_8$ ($M$\,=\,V,Ti,Mo,Nb,Ta; $A$\,=\,Ga,Ge,Al; $X$\,=\,S,Se,Te) 
\cite{Barz1973,Yaich1984,Pocha2000,Pocha2005,Geirhos2022Rev}, see Fig.\ \ref{fig:structure}. 
With one electron in a quasimolecular $t_2$ orbital, ideal $J_{\rm tet}$\,=3/2 moments forming 
an \textit{fcc} lattice have been claimed to be realized in $5d$ GaTa$_4$Se$_8$ 
\cite{Kim2014,Jeong2017direct}. 
Remarkably, a cluster Mott character has also been proposed, mainly based on band-structure 
calculations, for the $4d$ and even the $3d$ compounds, where smaller hopping competes 
with larger on-site Coulomb repulsion $U$ 
\cite{Johrendt1998,Pocha2000,AbdElmeguid2004,Kim2014,Nikolaev19,Lee19,Kim2020,Hozoi2020,Geirhos2022Rev,Petersen2023}. 
However, a direct experimental proof of quasimolecular electronic states in the lacunar 
spinels is still lacking. 
Such a quasimolecular character is particularly intriguing in the light of the 
complex phase diagrams of the lacunar spinels, which include multiple multiferroic 
and skyrmion-lattice phases with, e.g., N\'{e}el-type skyrmions carrying electric 
polarization \cite{Kezsmarki15,Ruff15,Bordacs17,Butykai22}, 
\mbox{(anti-)}ferroelectric states with peculiar domain architectures \cite{Ghara21,Geirhos21,Puntigam22}, 
and magnetism tied to polar domain walls \cite{Geirhos20}. 
The $5d$ Ta compounds host a pressure-induced insulator-to-metal transition followed by topological 
superconductivity \cite{AbdElmeguid2004,Park20,Deng2021,Jeong2021_RIXSpressure} 
and an avalanche-type dielectric breakdown of the Mott gap \cite{Guiot13}.

\begin{figure*}
	\centering
	\includegraphics[width=\textwidth]{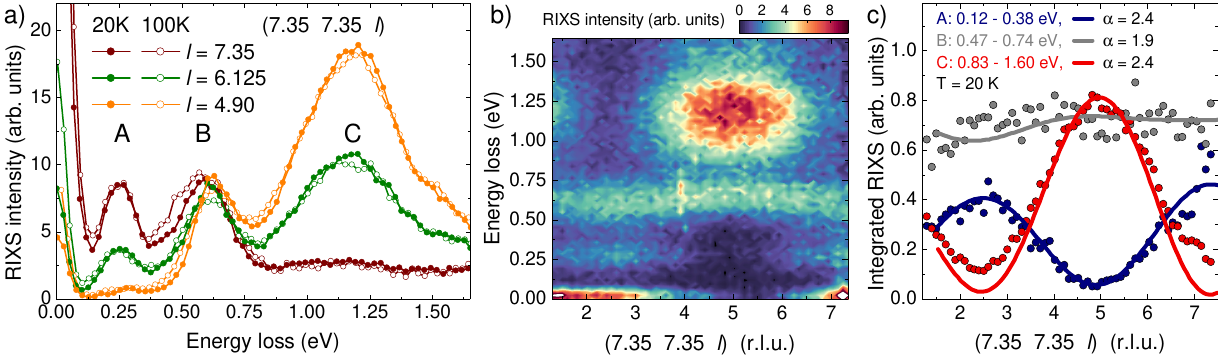}
	\caption{RIXS data of GaTa$_4$Se$_8$ along (7.35\,\,7.35\,\,$l$). 
	a) Spectra acquired at 20 and 100\,K show the three peaks A, B, and C.\@ 
	Changing $\mathbf{q}$ strongly affects the intensity. 
	b) Color map of the RIXS intensity at 20\,K.\@ Independent of the cluster modulation, 
	the elastic line is suppressed around $l$\,=\,5.4 due to a scattering angle 
	$2\theta$ close to 90$^\circ$. 
	c) Integrated intensity of peaks A, B and C.\@ Integration intervals are given 
	in the panel. Peaks A and C show dominant $\sin^2(\pi l/4.9)$ and 
	$\cos^2(\pi l/4.9)$ behavior, respectively. Solid lines: Results of the single-particle model.
	}
	\label{fig:spectra_map_intergral}
\end{figure*}

Here, we address the cluster wavefunction, which is the essential starting point for 
a comprehensive understanding of the lacunar spinels. 
We study GaTa$_4$Se$_8$ via resonant inelastic x-ray scattering (RIXS) at the Ta $L_3$ edge. 
RIXS directly probes the quasimolecular nature of, e.g., intra-$t_2^1$ excitations and pinpoints  
that the electrons are fully delocalized over a Ta$_4$ tetrahedron while correlations hardly 
affect the $t_2^1$ manifold. We find that the quasimolecular $J_{\rm tet}$\,=\,3/2 wavefunction 
deviates from the idealized case assumed previously \cite{Kim2014,Jeong2017direct}, 
since competing intracluster hopping terms mutually mix the corresponding bonding and 
antibonding orbitals. In GaTa$_4$Se$_8$, this mixing reduces the effective spin-orbit coupling 
constant $\zeta_{\rm eff}$ by roughly 1/3.  
Arising from strong hopping, the mixing is not a small perturbation 
and can be expected to affect the exchange interactions. Based on this mixing, 
the cluster wavefunction is sensitive to structural changes due to, e.g., 
external pressure or chemical substitution, which provides a new perspective on 
the entire family of lacunar spinels.

The delocalization of electrons over a cluster yields a characteristic modulation of the 
RIXS intensity $I(\mathbf{q})$ as function of the transferred momentum $\mathbf{q}$. 
This modulation reflects the dynamical structure factor $S(\mathbf{q},\omega)$ 
and reveals the character and symmetry of electronic states. 
For a dimer, RIXS can be described as an inelastic version of Young's double-slit experiment 
\cite{Ma1995interference}. The corresponding sinusoidal interference pattern has been observed 
recently in Ba$_3$CeIr$_2$O$_9$ and related dimer compounds 
\cite{Revelli2019_double_slit,Revelli_In,Magnaterra2022_Ti}. 
Stunningly, a sinusoidal intensity modulation has also been found in the Kitaev material 
Na$_2$IrO$_3$ where it unravels the bond-directional nearest-neighbor character of the 
magnetic excitations \cite{Revelli20,Magnaterra23}. 
Careful consideration of these interference effects is a prerequisite for the analysis of 
RIXS in cluster Mott insulators and provides a powerful tool to address the electronic 
states of GaTa$_4$Se$_8$.

Single crystals of GaTa$_4$Se$_8$ were grown by chemical vapor transport \cite{Winkler22}. 
At 300\,K, GaTa$_4$Se$_8$ shows the noncentro\-symmetric cubic space group $F\bar{4}3m$ 
with lattice constant $a$\,=\,10.382\,\AA\ \cite{Pocha2005}, see Fig.\ \ref{fig:structure}. 
The short intratetrahedral Ta-Ta distance $d$\,=\,3.0\,\AA\ suggests a quasimolecular character. 
The optical conductivity characterizes the lacunar spinels as narrow-gap insulators 
and reveals a Mott gap of 0.12\,eV in GaTa$_4$Se$_8$ \cite{Guiot13,TaPhuoc13,Reschke20}.
Experimental results for the magnetic moment per Ta$_4$ cluster yield 
0.7-1.2\,$\mu_{\rm B}$ \cite{Kawamoto16,Petersen2022_corr,Ishikawa2020,Pocha2005}. 
The magnetostructural transition at $T_{\rm ms}$\,=\,53\,K is accompanied by a strong drop 
in the magnetic susceptibility \cite{Ishikawa2020,JakobPhD2007,Kawamoto16}, but the 
crystal symmetry at low temperature is still under debate \cite{JakobPhD2007,Ishikawa2020,Geirhos2022Rev,Yang2022_bondordering}.
We first focus on cubic symmetry and then address the effect of distortions.

We measured RIXS at the Ta $L_3$ edge at Sector 27 at the 
Advanced Photon Source \cite{Shvydko_APS_MERIX}. 
The incident energy 9.879\,keV resonantly enhances excitations within 
the Ta $t_{2g}$ orbitals \cite{Jeong2017direct}.
We studied a (111) surface with the (110) and (001) directions in the horizontal scattering plane, 
using incident $\pi$ polarization. 
An energy resolution $\Delta E$\,=\,76\,meV was achieved using a Si(440) 
four-bounce monochromator and a R\,=\,2\,m Si(066) spherical diced crystal analyzer. 
We subtracted a background intensity that was determined by averaging over a range of negative 
energy loss. All spectra have been corrected for geometrical self-absorption \cite{Minola15}. 
We express $\mathbf{q}$ in reciprocal lattice units (r.l.u.). 
The $\mathbf{q}$ resolution equals $\Delta\mathbf{q}$\,=\,(0.1\,\,0.1\,\,0.3).

The RIXS spectra of GaTa$_4$Se$_8$ show three peaks A, B, and C at about 0.25, 0.62, and 1.2\,eV, 
see Fig.\ \ref{fig:spectra_map_intergral}a). The peak energies hardly show any dispersion 
but the intensity strongly depends on $\mathbf{q}$, in agreement with the data of 
Jeong \textit{et al}.\ \cite{Jeong2017direct}. 
This is a first indication of the local, quasimolecular character of the electronic 
states. For the peak assignment, we address the electronic structure of a single 
Ta$_4$ tetrahedron, starting with a non-interacting picture in the undistorted 
cubic case. Note that the RIXS data at 20\,K and 100\,K, i.e., 
above and below the structural transition at 53\,K, are very similar.

\begin{figure}[b]
	\centering
	\includegraphics[width=\columnwidth]{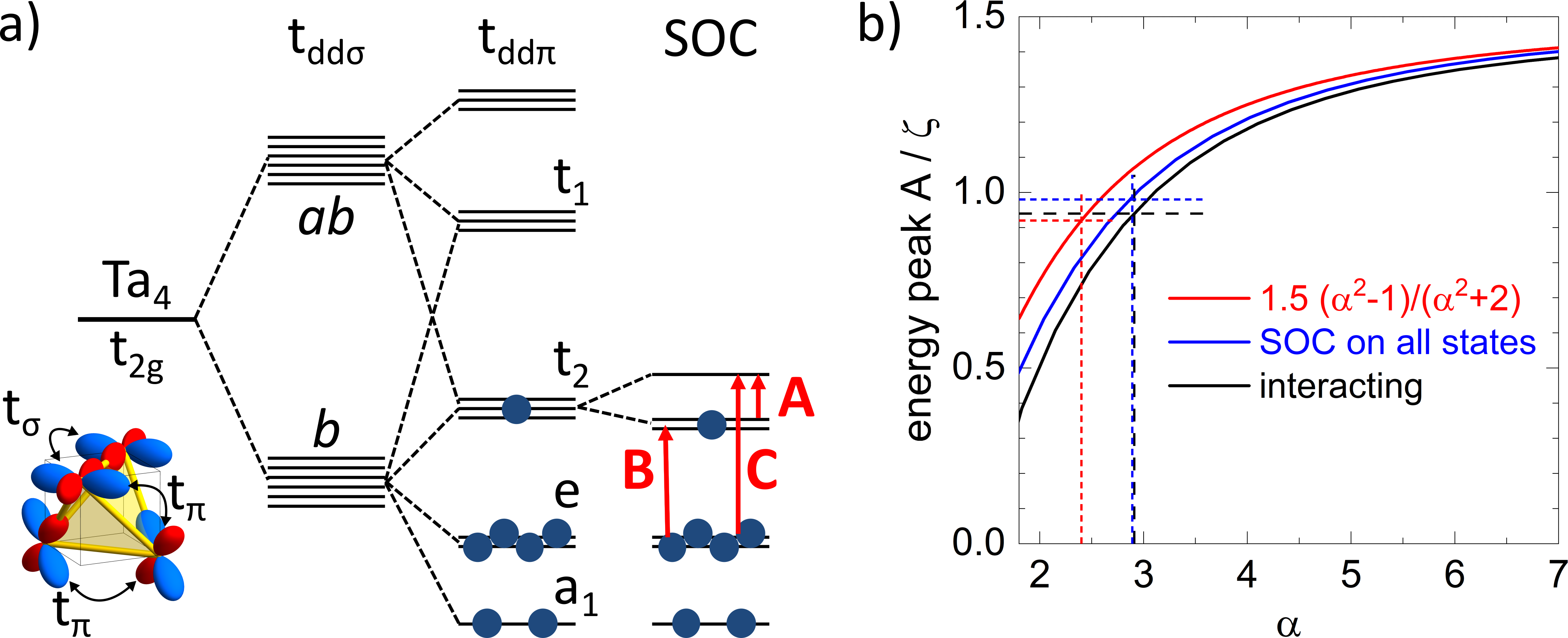}
	\caption{a) Single-particle energy levels of a Ta$_4$ tetrahedron. 
	Intra\-cluster hopping (see bottom left) yields quasimolecular orbitals and an $a_1^2 e^4 t_2^1$ ground state. 
	Because of $t_\pi$, the $t_2$ orbitals show contributions of bonding ($b$) 
	and antibonding ($ab$) states of $t_\sigma$. 
	Spin-orbit coupling within the $t_2^1$ states forms a $J_{\rm tet}$\,=\,3/2 ground state. 
	A, B, and C refer to the RIXS peaks, see Fig.\ \ref{fig:spectra_map_intergral}. 
	b) The admixture of antibonding character renormalizes the energy of peak A, see Eq.\ (\ref{eq:t2AB}). 
	Red (blue): single-particle result for spin-orbit coupling within $t_2$ (all) states. 
	Black: many-body cluster calculation using Quanty \cite{Haverkort_2016},
	see \textit{Supplementary Information} \cite{SI}.
	Dashed lines: value of $\alpha$ derived from the $\mathbf{q}$ dependence 
	and corresponding excitation energy.
}
\label{fig:hopping}
\end{figure}

Due to the large cubic crystal-field splitting 10\,$Dq$\,$\approx$\,3\,eV \cite{Jeong2017direct}, 
it is sufficient to consider the Ta $t_{2g}$ states. Direct $\sigma$-type hopping 
$t_\sigma$\,$\equiv$\,$t_{dd\sigma}$ of order 1\,eV \cite{Kim2014} yields bonding 
($b$) and antibonding ($ab$) states at $\pm t_\sigma$. 
Adding $\pi$-type hopping $t_\pi$\,$\equiv$\,$t_{dd\pi}$ results in 
the quasimolecular orbitals $a_1$, $e$, and $t_2$ at low energy and 
further states at high energy, see Fig.\ \ref{fig:hopping}a). 
With 7 electrons per Ta$_4$ cluster, the ground state shows fully occupied 
$a_1$ and $e$ orbitals plus a single electron in the $t_2$ states, $a_1^2 e^4 t_2^1$.  
The three $t_2$ orbitals are central to our discussion. Due to $t_\pi$, they are mixtures 
of bonding and antibonding states of $t_\sigma$, see Fig.\ \ref{fig:hopping}a). 
We will show the relevance of this mixture below but first follow the typical assumption 
where only the bonding ones are considered. This yields (cf.\ Fig.\ \ref{fig:structure}b)
\begin{equation}
\label{eq:xyB}
	xy_b  =  \left( xy_1 + xy_2 + xy_3 + xy_4 \right)/2  
\end{equation}
and equivalent for $yz_b$ and $xz_b$, where $i$\,=\,1-4 denotes the Ta sites. 
Projecting spin-orbit coupling $\zeta$ onto the subspace of these $t_2^1$ states 
yields a cluster Hamiltonian that is fully equivalent to the single-site case \cite{Jackeli09}, 
see \textit{Supplementary Information} \cite{SI}. It shows a $J_{\rm tet}$\,=\,3/2 ground 
state and a $J_{\rm tet}$\,=\,1/2 excited state at $1.5 \zeta$ \cite{Kim2014}. 
The latter, the so-called spin-orbit exciton, corresponds to peak A, 
while peaks B and C in this non-interacting scenario are assigned to excitations from 
$e$ to $J_{\rm tet}$ states, i.e., from $e^4t_2^1$ 
to $e^3 t_2^2$, see Fig.\ \ref{fig:hopping}a).

This peak assignment is supported by quantum chemistry calculations \cite{Petersen2022_corr} 
and is confirmed by the characteristic $\mathbf{q}$ dependence of the RIXS intensity.  
Figure \ref{fig:spectra_map_intergral}b) is a color plot for energies up to 1.6\,eV, 
while Fig.\ \ref{fig:spectra_map_intergral}c) shows the integrated RIXS intensity of peaks A, B, 
and C for $\mathbf{q}$ along (7.35\,\,7.35\,\,$l$) together with results of a single-particle 
calculation (see below). 
Peak B hardly depends on $\mathbf{q}$, while A and C show a pronounced 
$\sin^2(\pi l/4.9)$ and $\cos^2(\pi l/4.9)$ behavior, respectively, reflecting the different 
symmetries of the corresponding states. 
The period $l_0$\,=\,4.9 points to a Ta-Ta distance of $a/4.9$\,$\approx$\,2.12\,\AA\ 
that agrees with the $c$-axis projection $d/\sqrt{2}$\,$\approx$\,2.12\,\AA\ 
of the intratetrahedral Ta-Ta distance $d$. 
Spectra for $l$\,=\,$l_0$ and $l$\,=\,$1.5l_0$\,=\,7.35 correspond to extrema of the 
intensity modulation, cf.\ Fig.\ \ref{fig:spectra_map_intergral}a).

\begin{figure}[t]
	\centering
	\includegraphics[width=\columnwidth]{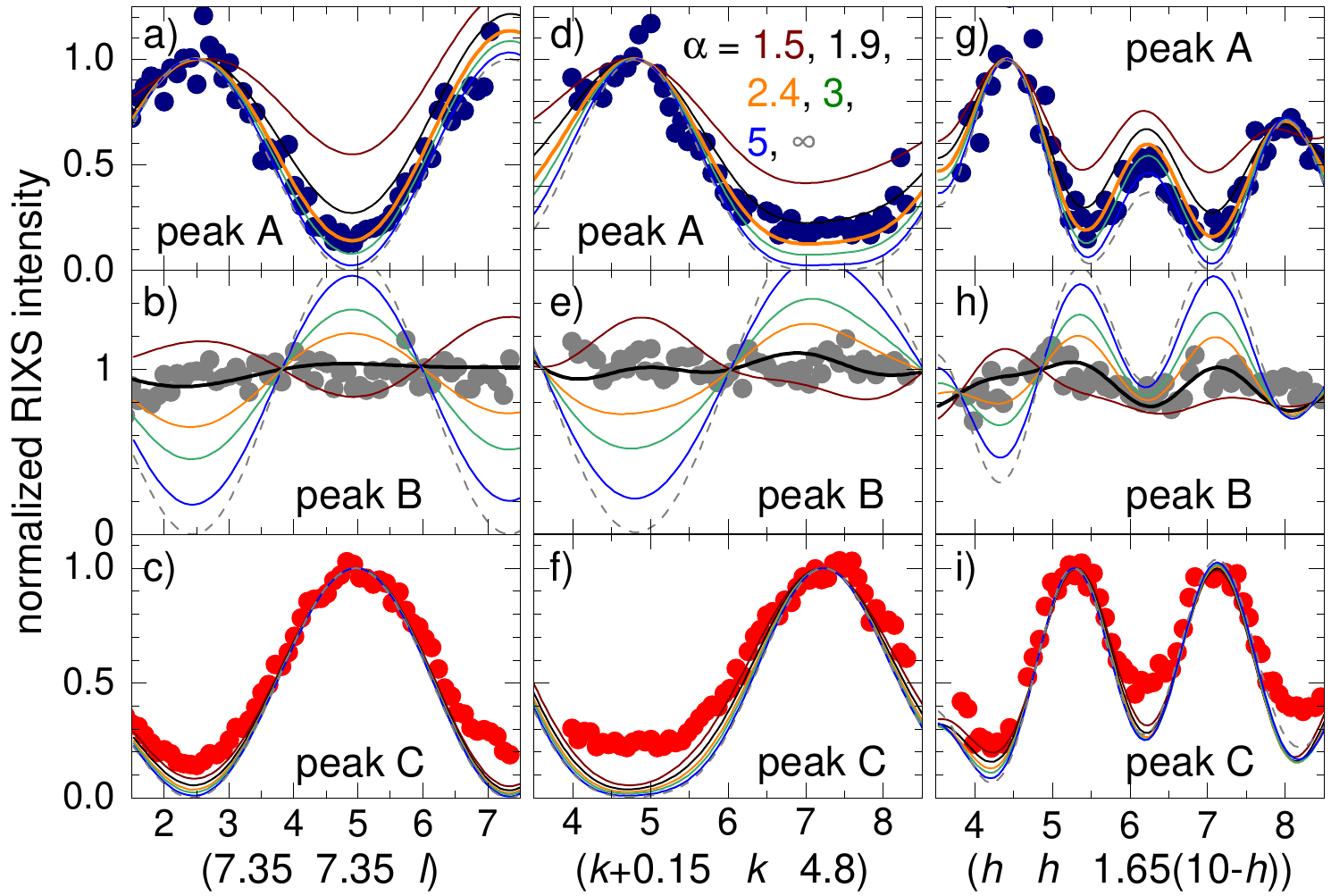}
	\caption{Normalized RIXS intensity (symbols) along three $\mathbf{q}$ directions 
	for peaks A, B and C with integration ranges as in Fig.\ \ref{fig:spectra_map_intergral}. 
	Lines: Results of the single-particle model for spin-orbit coupling within the $t_2$ states. 
	Note that $\alpha$ is the only free parameter.
	For peaks A and B, best agreement is obtained for $\alpha$\,=\,$2.4\pm 0.3$ and 1.8-1.9, 
	respectively.
	Dashed: Result for $\alpha$\,=\,$\infty$, neglecting antibonding states. 
	}
	\label{fig:model}
\end{figure}

The dominant $\sin^2(\pi l/4.9)$ behavior of peak A is a clear fingerprint of the 
quasimolecular intra-$t_2^1$ spin-orbit exciton. 
In general, the RIXS intensity for an excitation from the ground state 
$|0\rangle$ to a final state $|f\rangle$ is described by
\cite{Ament2011_fast,Ament2011_RIXSrev}
\begin{equation}
	I_f(\mathbf{q},\omega) = \Big| \langle f |\, \sum_{\gamma} e^{i\mathbf{q}\cdot\mathbf{R}_\gamma}\, D^\dagger_{\gamma} D _{\gamma} \,| 0\rangle \Big|^2 \delta(\hbar \omega - E_f)
	\label{eq:RIXS_intensity_general}
\end{equation}
where $E_f$ denotes the excitation energy and $D_\gamma$ $(D^\dagger_\gamma)$ is the local dipole 
operator for resonant scattering at the Ta site $\mathbf{R}_\gamma$. 
This coherent sum of local scattering processes is 
running over all $\mathbf{R}_\gamma$ from which the final state $|f\rangle$ can be reached. 
For the quasimolecular states in GaTa$_4$Se$_8$, this refers to the four Ta sites of a tetrahedron.
For $\mathbf{q}$ along (7.35\,\,7.35\,\,$l$), the physics is particularly simple if we stick 
to the contribution of bonding states to the quasimolecular $t_2$ orbitals, 
see Fig.\ \ref{fig:hopping}a), 
i.e., we employ Eq.\ (\ref{eq:xyB}) and the associated $J_{\rm tet}$ states for 
spin-orbit coupling within the $t_2$ states, as discussed above. 
In this case, $I_{B/C}(l) \! \propto \! \cos^2(\pi l/4.9)$ 
for all excitations from $e$ to $t_2$, while only the spin-orbit exciton is 
expected to show $I_A(l) \! \propto \! \sin^2(\pi l/4.9)$, 
see dashed lines in Fig.\ \ref{fig:model}a)-c). 
This firmly supports our interpretation of peak A.

Our central goal is to determine the cluster wavefunction. 
Thus far, we considered only the bonding contributions to the $t_2$ orbitals, 
see Eq.\ (\ref{eq:xyB}), a common approximation \cite{Kim2014,Jeong2017direct}.
In this simple bonding picture, $I_C(l)$ describes the overall behavior of peak C but
$I_B(l)$ does not explain the nearly $\mathbf{q}$-independent intensity of peak B.\@ 
Furthermore, this approximation predicts the spin-orbit exciton at 1.5$\zeta$, 
as for a single site, which is hard to reconcile with the energy of peak A at 0.25\,eV.\@
The equivalent excitation for weakly interacting Ta $5d^1$ ions has been observed in RIXS 
on Rb$_2$TaCl$_6$ at 0.4\,eV \cite{Ishikawa2019}, resulting in $\zeta$\,$\approx$\,0.27\,eV.\@ 
Compared to 0.4\,eV, the energy of peak A is about 40\,\% smaller.  
As shown below, these critical issues are resolved by considering the admixture 
of antibonding character to the $t_2$ orbitals. 
With the intracluster hoppings $t_\sigma$ and $t_\pi$, the eigenstate $t_2(xy)$ 
of the hopping Hamiltonian reads
\begin{equation}
	\label{eq:t2AB}
	t_2(xy) = \left[xy_b - (yz_{ab} - xz_{ab})/\alpha 
	\right]/\sqrt{1+(2/\alpha^2)}
\end{equation}
with the antibonding orbitals $yz_{ab}$\,=\,$(yz_1 - yz_2 + yz_3 - yz_4)/2$ 
and $xz_{ab}$\,=\,$(xz_1 - xz_2 - xz_3 + xz_4)/2$. 
The approximation of Eq.\ (\ref{eq:xyB}) corresponds to $\alpha$\,=\,$\infty$. 
The mixing coefficient reads 
\begin{equation}
	\label{eq:alpha}
	\alpha = |t_\sigma/t_\pi| - 3/2 +\sqrt{(|t_\sigma/t_\pi| - 3/2)^2+2} \, .
\end{equation}
Projecting spin-orbit coupling onto this $t_2^1$ subspace yields the 
same cluster Hamiltonian as above \cite{SI} but with renormalized coupling constant 
$\zeta_{\rm eff}$\,=\,$\zeta \cdot  (\alpha^2 - 1) \, / \, (\alpha^2 + 2)$.
Accordingly, the peak assignment of Fig.\ \ref{fig:hopping}a) is still valid but 
$\alpha$ renormalizes in particular the energy of peak A, cf.\ Fig.\ \ref{fig:hopping}b), 
and changes the character of the quasimolecular $J_{\rm tet}$ states. Using Harrison's 
empirical $d$ dependence of the Slater-Koster parameters \cite{Harrison}, we find 
$|t_\sigma/t_\pi|$\,=\,$1.5 |V_{dd\sigma}/V_{dd\pi}|$\,$\approx$\,2.8. 
This yields a first estimate $\alpha$\,$\approx$\,3.2.
Taking into account hopping $t_{\rm Se}$ via the Se ligands 
reduces $\alpha$, for instance to $\alpha$\,$\approx$\,2 for 
$t_{\rm Se}$\,$\approx$\,$t_\pi$.
The $t_2(xy)$ orbital for $\alpha$\,=\,2 is depicted in Fig.\ \ref{fig:structure}c).

Experimentally, the $\mathbf{q}$-dependent RIXS intensity is the ideal tool to determine 
the mixing coefficient $\alpha$. Via the matrix elements in 
Eq.\ (\ref{eq:RIXS_intensity_general}), RIXS is directly sensitive to the quasimolecular 
wavefunction and hence to the admixture of antibonding orbitals. 
We calculated the RIXS response in the single-particle picture for spin-orbit coupling within 
the $t_2$ states, taking into account polarization selection rules. 
Results for the normalized RIXS intensities of peaks A, B, and C along (7.35\,\,7.35\,\,$l$) 
and two further $\mathbf{q}$ directions are plotted in Fig.\ \ref{fig:model}. 
Along ($k$+0.15\,\,$k$\,\,4.8), the dominant term for peak A is $\cos^4(\pi k/4.9)$ 
while a more complex behavior is observed along ($h$\,\,$h$\,\,1.65(10-$h$)). 
The single-particle picture captures the behavior of all three peaks surprisingly well. 
We emphasize that $\alpha$ is the only free parameter in Fig.\ \ref{fig:model},  
reflecting the dependence of the wavefunction on $t_\sigma$ and $t_\pi$.
These results unambiguously establish the quasimolecular cluster-Mott character of GaTa$_4$Se$_8$ 
and that the admixture of antibonding character is sizable, i.e., $1/\alpha$ is not small.

The single-particle picture is expected to work particularly well for the intra-$t_2^1$ excitation 
of the spin-orbit exciton, peak A.\@ Peaks B and C with $e^3 t_2^2$ final states will be more 
sensitive to interactions. Peak C is the least sensitive to $\alpha$. 
The single-particle model reproduces the overall $\mathbf{q}$ dependence but fails to describe 
the minima quantitatively. To some extent, this may reflect a possible background contribution 
of excitations across the Mott gap at high energies.
In contrast, peak B is highly sensitive to $\alpha$. Its nearly constant behavior as function 
of $\mathbf{q}$ is reproduced in a narrow window $\alpha$\,=\,1.8-1.9. 
For peak A, excellent agreement is found for $\alpha$\,=\,$2.4\pm 0.3$.

These results for $\alpha$ fall in the range predicted above based on Harrison's 
rules. The precise value depends on details of the model concerning 
the range of spin-orbit coupling, distortions, sub-leading hopping terms, and correlations. 
Above, spin-orbit coupling was projected onto $t_2$ orbitals only. If we instead consider 
all orbitals, in particular including $e$ and $t_1$ (see Fig.\ \ref{fig:hopping}a), 
peak A is best described for $\alpha$\,=\,$2.9\pm 0.4$. Furthermore,
we discussed regular tetrahedra but the symmetry is lower than cubic 
below $T_{ms}$\,=\,53\,K.\@ 
Recent x-ray and neutron results on the pair distribution function 
\cite{Yang2022_bondordering} point to dynamical local distortions up to temperatures 
far above $T_{ms}$. 
For trigonally distorted tetrahedra, we find that RIXS is sensitive to the distortion 
if a single orientation can be studied while the average over different orientations 
is very close to the cubic case, see \textit{Supplementary Information} \cite{SI}. 
The latter applies to both 20 and 100\,K and validates our approach. 
However, a distortion of the tetrahedra with less than trigonal symmetry 
affects the value of $\alpha$ for which peak B is nearly independent of $\mathbf{q}$, 
see \textit{Supplementary Information} \cite{SI}.  
Note that different results were reported for the crystal symmetry at low temperature 
\cite{JakobPhD2007,Ishikawa2020,Geirhos2022Rev,Yang2022_bondordering}, 
impeding an even more precise determination of $\alpha$ at present.

In a cluster Mott insulator, electron-electron interactions suppress intercluster charge 
transport. Within a cluster, correlations compete with dominant hopping that delocalizes 
the electrons in quasimolecular orbitals. To study the effect on the RIXS response, we performed 
many-body calculations for a single tetrahedron using Quanty \cite{Haverkort_2016}, 
see \textit{Supplementary Information} \cite{SI}.  
Interactions yield a fanning out of the $a_1^2 e^3 t_2^2$ energy levels that is relevant to 
explain the width of peak C and the energies of B and C, supporting our peak assignment. 
For peak A, we find that electron-electron interactions have only a minor effect 
on both the energy and the $\mathbf{q}$-dependent intensity, 
in particular for comparison with the case where spin-orbit coupling is 
not restricted to $t_2$, see \textit{Supplementary Information} \cite{SI}.  
The many-body calculations thus support the overall picture of the single-particle model.

The renormalized energy of the spin-orbit exciton, peak A, provides an independent means 
to test our results for $\alpha$. Figure \ref{fig:hopping}b) shows the single-particle result 
for spin-orbit coupling within $t_2$, 
$E_{\rm SO}$\,= $1.5\zeta_{\rm eff}$\,=\,$1.5\zeta\,(\alpha^2\!-\!1)/(\alpha^2\!+\! 2)$. 
For comparison, the excitation energy is also given for spin-orbit coupling acting on all 
states and for the many-body cluster calculation. For the latter we change $\alpha$ by 
changing $t_\sigma$, cf.\ Eq.\ (\ref{eq:alpha}), with all other parameters fixed. 
The overall behavior is very similar. For all three cases, the dashed lines denote the value 
of $\alpha$ that best describes the $\mathbf{q}$ dependence. This yields an excitation energy 
of  0.9-1.0\,$\zeta$ and hence $\zeta$\,$\approx$\,0.27-0.30\,eV, in very good agreement with 
both quantum chemistry calculations \cite{Petersen2022_corr} and the value 0.27\,eV reported 
for $5d^1$ Rb$_2$TaCl$_6$ \cite{Ishikawa2019}.

In conclusion, our results establish GaTa$_4$Se$_8$ as a fascinating example of a cluster 
Mott insulator. The valence electrons are fully delocalized over a Ta$_4$ tetrahedron,
while intercluster charge fluctuations are suppressed.
A thorough analysis of the modulated RIXS intensity $I(\mathbf{q})$
reveals the quasimolecular wavefunction, which is the essential starting point 
for exploring the physics of cluster Mott insulators. 
The spin-orbit exciton, an excitation within the $t_2^1$ manifold, is particularly well 
described in a single-particle scenario that is coined by competing hopping terms, 
$t_\sigma$ and $t_\pi$.
This competition shapes the wavefunction, renormalizes the effective spin-orbit coupling 
constant by roughly 1/3, and hence affects the nature of the quasimolecular magnetic moment. 
We expect that this is decisive for intercluster exchange coupling, calling for 
future theoretical investigations. 
In general, the mixing coefficient $\alpha$ also depends on $t_{dd\delta}$ and 
on the indirect hopping via ligands. Therefore, it is reasonable to assume that $\alpha$ 
can be tuned in the lacunar spinels by external pressure and chemical substitution, 
and one may speculate that even temperature may tip the balance in certain cases. 
Our results on the quasimolecular character, the particular role of antibonding 
states, and the tunability of the wavefunction are relevant 
for many of the open questions on the large family of lacunar spinels.

\begin{acknowledgments}
We thank D. I. Khomskii and J. van den Brink for fruitful discussions and 
A. Revelli for experimental support. 
We gratefully acknowledge the Advanced Photon Source (APS) for providing beam time 
and technical support. APS is a U.S. Department of Energy (DOE) 
Office of Science user facility operated for the DOE Office of Science by 
Argonne National Laboratory under Contract No.\ DE-AC02-06CH11357. 
Furthermore, we acknowledge funding from the Deutsche Forschungs\-gemeinschaft 
(DFG, German Research Foundation) via Project numbers 277146847 
(CRC 1238, projects B03, C03), 
492547816 (TRR 360), 
and 437124857 (KE 2370/3-1). 
V.T.\@ acknowledges the support via project ANCD 20.80009.5007.19 (Moldova).
M.H.\@ acknowledges partial funding by the Knut and Alice Wallenberg Foundation 
as part of the Wallenberg Academy Fellows project.
\end{acknowledgments}

\newpage
~
\newpage

\section*{Supplementary Information}

\section{Spin-orbit coupling on the cluster}

Before addressing the tetrahedral cluster, we start by considering a single site in cubic symmetry with 
the three $t_{2g}$ orbitals $X$\,=\,$yz$, $Y$\,=\,$zx$, and $Z$\,=\,$xy$. 
Applying spin-orbit coupling $H_{\rm SOC}$\,=\,$\zeta \,\vec{l}\,\vec{s}$ to the 
single electron configuration $t_{2g}^1$ 
couples $|X\uparrow\rangle$, $|Y\uparrow\rangle$, and  $|Z\downarrow\rangle$ 
as well as $|X\downarrow\rangle$, $|Y\downarrow\rangle$, and  $|Z\uparrow\rangle$ \cite{Tanabe}, 
which is described by the two Hamiltonians
\begin{equation}
	H_{X\uparrow,Y\uparrow,Z\downarrow}  = \left( \begin{array}{ccc}
		0  &  i & -1  \\ 
		-i  &  0 & i  \\ 
		-1  & -i & 0
	\end{array} \right) \, \frac{\zeta}{2}
\end{equation}
\begin{equation}
	H_{X\downarrow,Y\downarrow,Z\uparrow}  = \left( \begin{array}{ccc}
		0  & -i & 1  \\ 
		i  &  0 & i  \\ 
		1  & -i & 0
	\end{array} \right) \, \frac{\zeta}{2}
\end{equation}
where the indices denote the basis states. 
Altogether, this yields a $j$\,=\,3/2 ground state and a $j$\,=\,1/2 excited state at 3/2\,$\zeta$.

For the four Ta sites of a tetrahedron, we consider hopping $t_\sigma$ and $t_\pi$ between the 
$4 \times 3$\,$t_{2g}$ orbitals as well as spin-orbit coupling on each of the four sites. 
Then we transform to the quasimolecular basis states of the hopping Hamiltonian, see Fig.\ 3a) 
of the main text. For the single-particle model, we project spin-orbit coupling to the 
quasimolecular $t_2$ states. 	
(For comparison, we also consider the case of spin-orbit coupling acting on all states. 
These exceptions are clearly marked in the main text.)
This is a reasonable approximation as long as the energy difference between the quasimolecular 
states is large. In spirit, it is analogous to neglecting the $e_g$ orbitals in the discussion of 
a single site. We find a Hamiltonian that is equivalent to the single-site case, 
\begin{eqnarray}
	\label{eq:HSOC}
	H_{X_{\rm tet}\uparrow,Y_{\rm tet}\uparrow,Z_{\rm tet}\downarrow}  & = & \left( \begin{array}{ccc}
		0  &  i & -1  \\ 
		-i  &  0 & i  \\ 
		-1  & -i & 0
	\end{array} \right) \, \frac{\zeta}{2}\,\, \frac{\alpha^2-1}{\alpha^2 +2}
	\\
	H_{X_{\rm tet}\downarrow,Y_{\rm tet}\downarrow,Z_{\rm tet}\uparrow}  & = & \left( \begin{array}{ccc}
		0  & -i & 1  \\ 
		i  &  0 & i  \\ 
		1  & -i & 0
	\end{array} \right) \, \frac{\zeta}{2}\,\, \frac{\alpha^2-1}{\alpha^2 +2}
\end{eqnarray}
where $Z_{\rm tet}$\,=\,$t_2(xy)$ is defined in Eq.\ (3) of the main text. 
Equivalent expressions hold for $X_{\rm tet}$ and $Y_{\rm tet}$. 
Compared to a single site, the energy scale is renormalized by $\alpha$, see Fig.\ 3b). 
The normalization factor approaches 1 in the limit $\alpha$\,=\,$\infty$.

\begin{figure}[b]
	\centering
	\includegraphics[width=0.8\columnwidth]{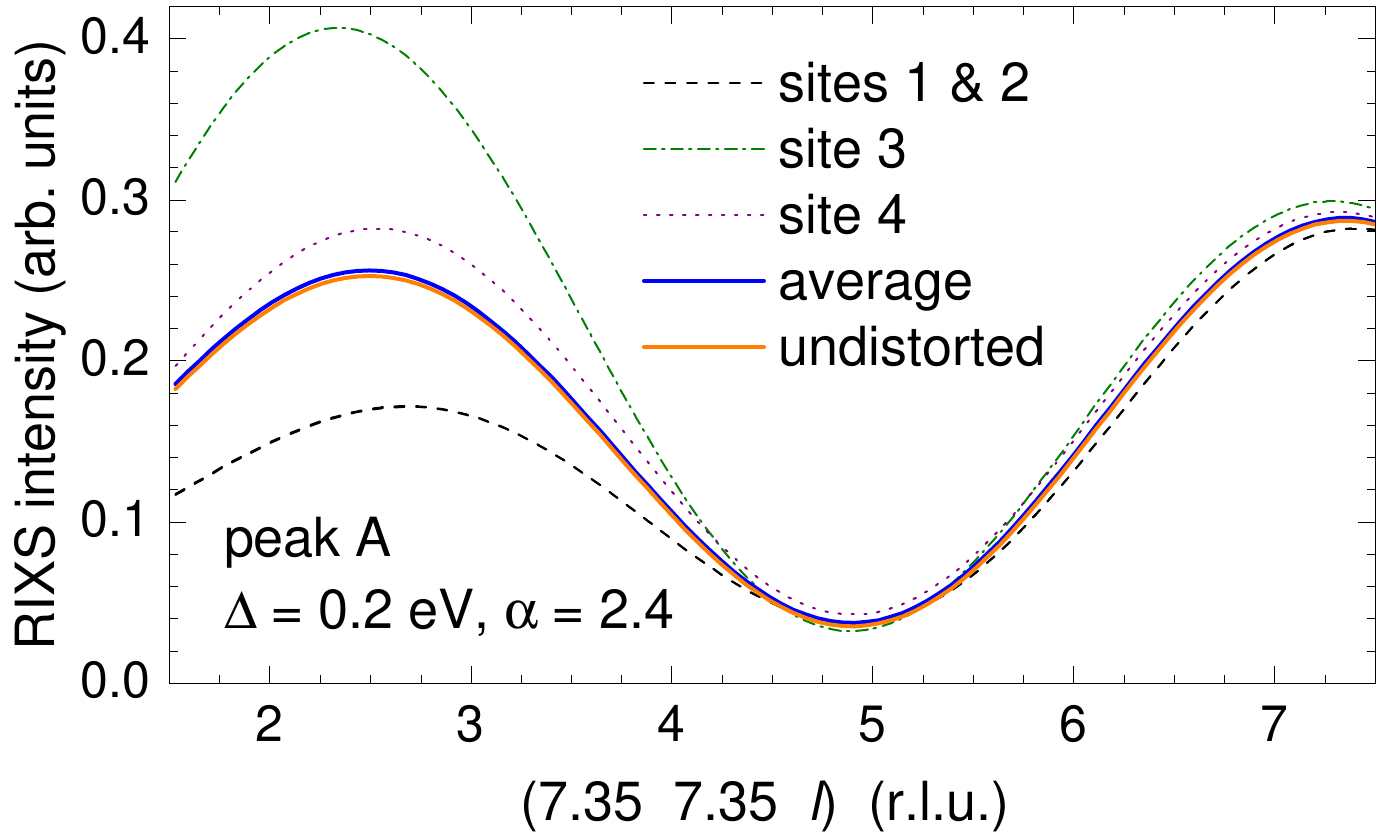}
	\caption{RIXS intensity of peak A for $\mathbf{q}$ along (7.35\,\,7.35\,\,$l$) for a 
		distorted tetrahedron with $\Delta$\,=\,0.2\,eV and $\alpha$\,=\,2.4. 
		For this $\mathbf{q}$ direction, the results for distortions with either site 1 or 2 as 
		unique site are identical. 
		The average over the four different orientations of the distortion (solid blue line) 
		is hard to distinguish from the undistorted case (solid orange line).
	}
	\label{fig:distorted}
\end{figure}

\begin{figure}[tb]
	\centering
	\includegraphics[width=0.9\columnwidth]{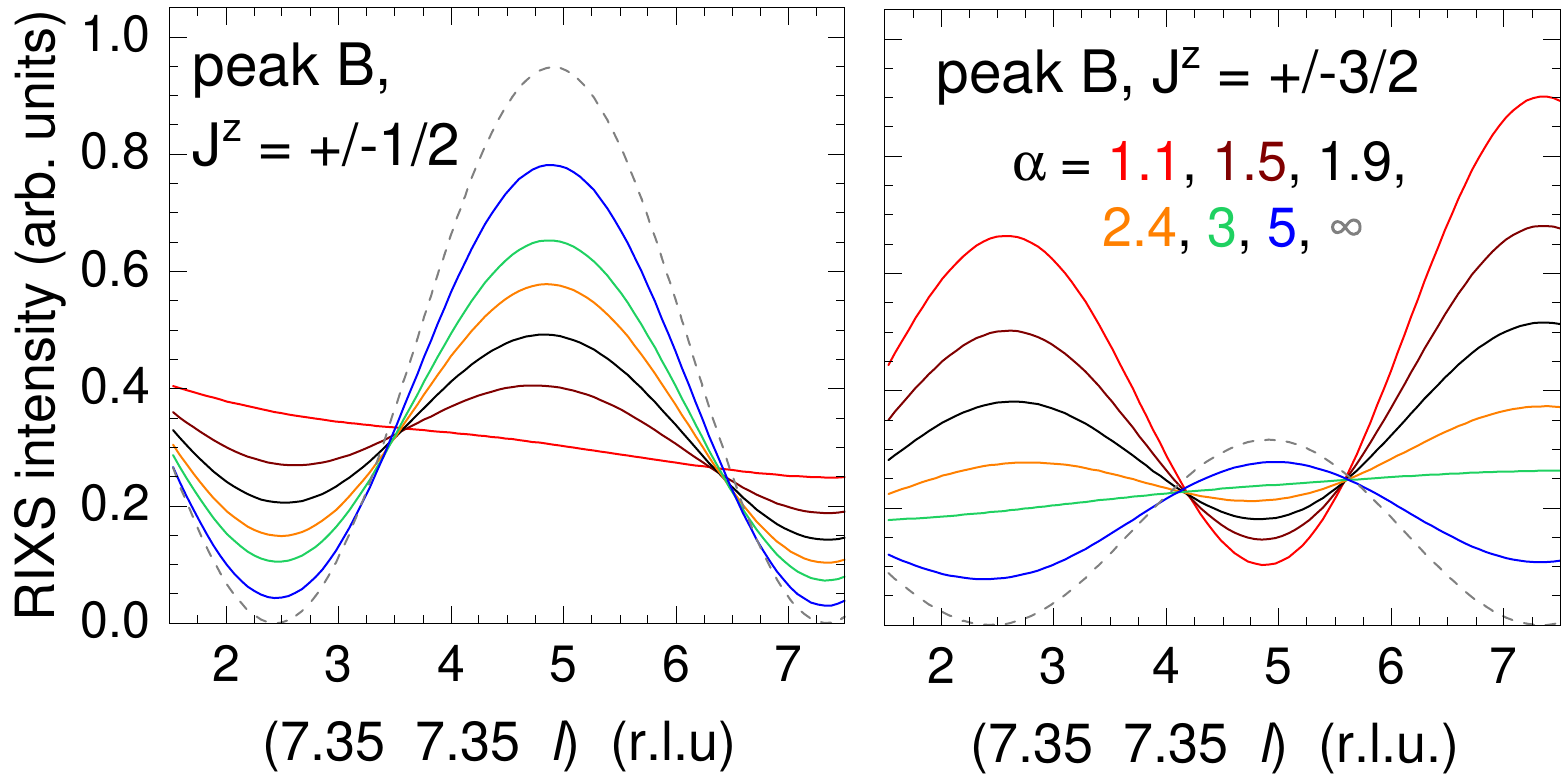}
	\caption{The $\alpha$ dependence of the RIXS intensities of the two different 
		contributions to peak B.\@ 
		Left (right): Excitations from an $e$ orbital to $J_{\rm tet}$\,=\,3/2 with $J^z$\,=\,$\pm$1/2 ($\pm$3/2). 		
	}
	\label{fig:distortedB}
\end{figure}

\section{Effect of non-cubic distortions}

In cubic symmetry, all Ta-Ta edges of the regular tetrahedron are of equal length $d$\,=\,3.0\,\AA. 
However, the symmetry is lower than cubic below $T_{ms}$\,=\,53\,K.\@ 
Furthermore, recent x-ray and neutron results on the pair distribution function 
\cite{Yang2022_bondordering} point to dynamical local distortions up to temperatures 
far above $T_{ms}$. These dynamical distortions are of similar size as the static 
distortions at low temperature. RIXS averages over different orientations of the dynamical distortions 
above $T_{ms}$ and different domains below $T_{ms}$. Even within a single domain, different tetrahedra 
within the unit cell show different orientations \cite{Yang2022_bondordering}. 
For a single tetrahedron, we approximate the distorted case by three short and three long bonds 
with $3.0 \pm 0.04$\,\AA, where the long bonds meet at one Ta site \cite{Yang2022_bondordering}. 
We implement such a trigonal distortion of the tetrahedron via the size of $\sigma$-type hopping, 
for which we assume $t_\sigma \mp \Delta/2$. 
With $t_\sigma \propto 1/d^5$ \cite{Harrison}, we estimate $\Delta/t_\sigma$\,$\approx$\,0.13. 
Below we use $\Delta$\,=\,0.2\,eV, which somewhat overestimates the effect of a trigonal distortion.

\begin{figure}[b]
	\centering
	\includegraphics[width=0.9\columnwidth]{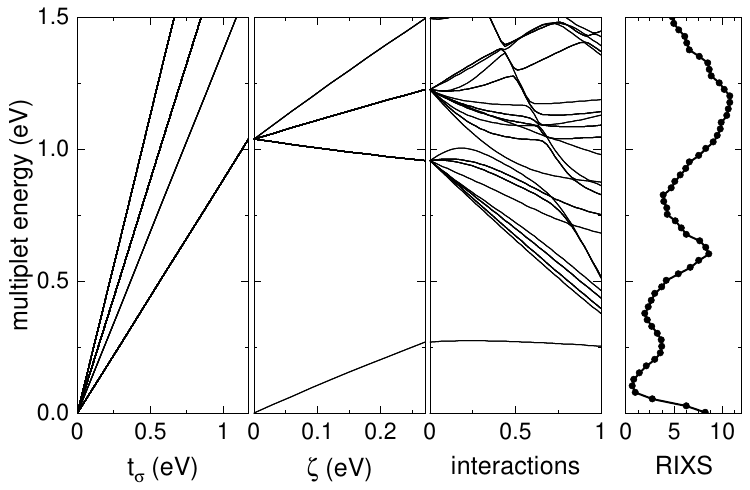}
	\caption{Excitation energies of a single Ta$_4$ tetrahedron (left) compared to a RIXS 
		spectrum (right). 
		The plot focuses on the low-energy sector, in particular on intra-$t_2$ 
		and $e$-to-$t_2$ excitations. 
		First panel: effect of hopping $t_\sigma$ for fixed 
		$|t_\sigma/t_\pi|$\,$\approx$\,2.6, i.e., $\alpha$\,=\,2.9. Beyond the $a_1^2 e^4 t_2^1$ 
		ground state manifold at zero energy it shows, with increasing energy, the states 
		$a_1^2 e^3 t_2^2$, $a_1^2 e^4 t_2^0 t_1^1$, $a_1^2 e^2 t_2^3$, and $a_1^2 e^3 t_2^1 t_1^1$.  
		Second panel: finite $\zeta$ lifts the degeneracy of the $a_1^2 e^3 t_2^2$ states 
		at about 1\,eV while the $a_1^2 e^4 t_2^1$ ground state is split into $J_{\rm tet}$\,=\,3/2 
		and 1/2 with an excitation energy of about $\zeta$. 
		Third panel: Electron-electron interactions are necessary to correctly describe the 
		excitation energies of peaks B and C.\@  
		The scale 0 to 1 encodes the linear increase from $J_H$\,=\,$U$\,=\,0 to 
		$J_H$\,=\,0.4\,eV and $U$\,=\,1.75\,eV.
	}
	\label{fig:Quanty_energy_levels}
\end{figure}

As in the cubic case, we employ the single-particle picture and first calculate the eigenstates 
of the hopping Hamiltonian for a single, distorted tetrahedron. Then we apply spin-orbit coupling 
within the $t_2^1$ states. The distortion splits the $J_{\rm tet}$\,=\,3/2 quartet into two 
doublets, in close analogy to the single-site case. We calculate RIXS for the four equivalent 
orientations of the distortion and finally average over the four curves. 
Figure \ref{fig:distorted} shows the result for peak A along 
(7.35\,\,7.35\,\,$l$) for $\Delta$\,=\,0.2\,eV and $\alpha$\,=\,2.4. 
We find clear differences for different orientations of the distortion. 
RIXS thus would be sensitive to the distortion if a single orientation of the distortion could 
be measured. 
However, the average over the four distorted results is very close to the RIXS response 
of the undistorted case. 
This is a plausible explanation for our RIXS data being very similar at 20\,K and 100\,K 
and validates the description of the data in terms of a cubic model.

Peak B corresponds to excitations from $e$ to $J_{\rm tet}$\,=\,3/2 states. 
The $J_{\rm tet}$\,=\,3/2 quartet contains states with $J^z$\,=\,$\pm 1/2$ and $J^z$\,=\,$\pm 3/2$, 
and these show a different $\mathbf{q}$ dependence, see Fig.\ \ref{fig:distortedB}. 
In particular a nearly $\mathbf{q}$-independent behavior is observed for $\alpha$\,$\approx$\,1.1 
for $J^z$\,=\,$\pm 1/2$ but for $\alpha$\,$\approx$\,3 for $J^z$\,=\,$\pm 3/2$. 
The trigonal distortion considered above mixes these two states with equal weight, 
and hence has little effect on the $\mathbf{q}$ dependence. 
However, a distortion of lower symmetry 
may cause preferential occupation of, e.g., $J^z$\,=\,$\pm 1/2$ in the ground state, 
which will partially block the corresponding contribution. 
In this case, a $\mathbf{q}$-independent behavior is expected for a larger value of $\alpha$ 
up to $\alpha$\,=\,2.2. It is plausible that such a low-symmetry distortion may describe 
the behavior of peaks A and B for the same value of $\alpha$.

\section{Electron-electron interactions}

Beyond the calculations within the single-particle model, we also performed many-body 
cluster calculations using Quanty \cite{Haverkort_2016}. 
For the seven $t_{2g}$ electrons on a Ta$_4$ tetrahedron, we consider the model Hamiltonian
$\mathrm{H} = \mathrm{H}_{t} + \mathrm{H}_{\zeta} + \mathrm{H}_{e-e}$ 
that takes into account hopping $t_\sigma$ and $t_\pi$, spin-orbit coupling $\zeta$, 
and electron-electron interactions, i.e., on-site Coulomb repulsion $U$ and 
Hund's coupling $J_H$. We first diagonalize $H_t$ and then consider 
the lower nine ($a_1$, $e$, $t_2$, and $t_1$) out of 12 quasimolecular orbitals, 
see Fig.\ 3a) of the main text.

\begin{figure}[bt]
	\centering
	\includegraphics[width=0.8\columnwidth]{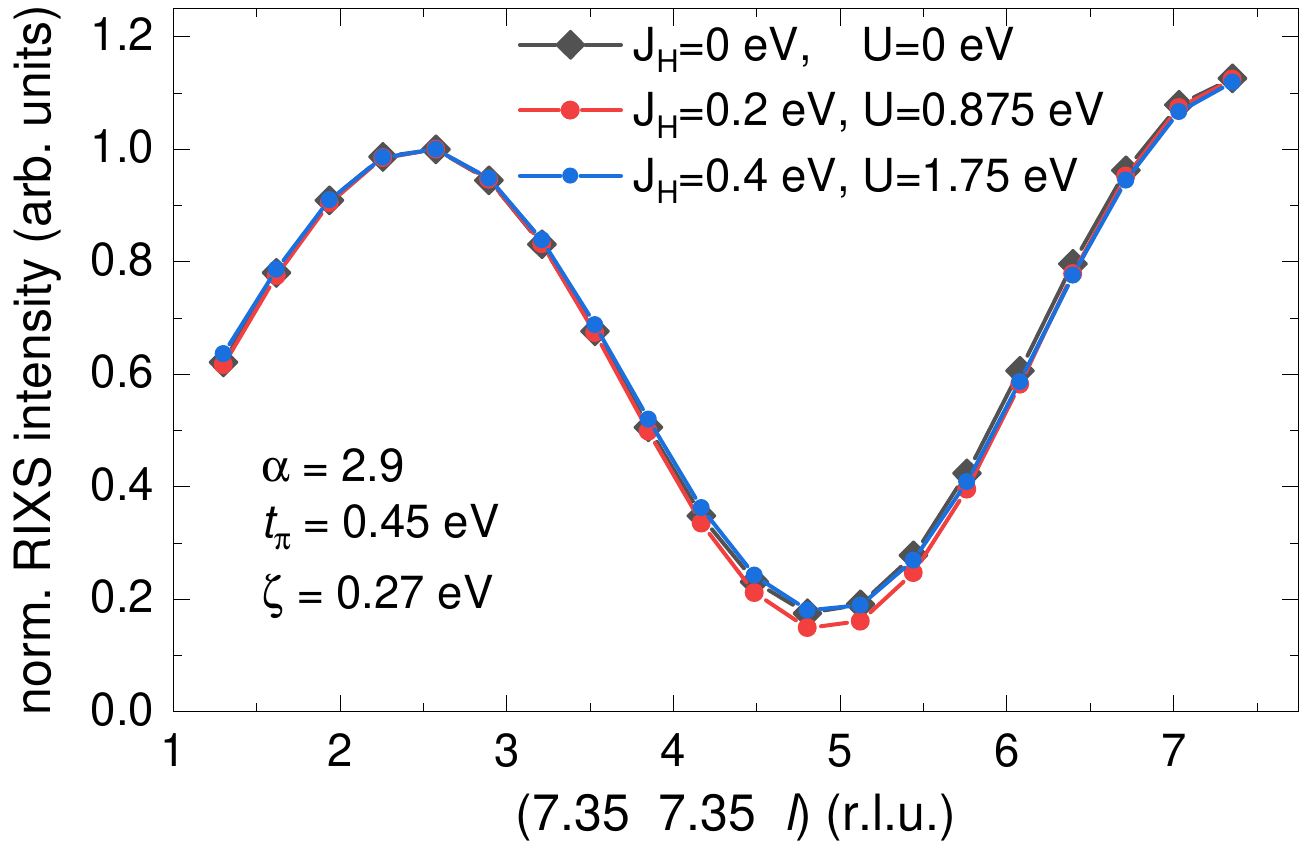}
	\caption{Effect of electron-electron interactions on the $\mathbf{q}$ 
		dependence of the normalized RIXS intensity of peak A along (7.35\,\,7.35\,\,$l$) for $\alpha$\,=\,2.9.	
	}
	\label{fig:Quanty_spectra_map_intergral}
\end{figure}

Taking into account electron-electron interactions is in particular necessary 
in order to correctly reproduce the excitation energies of peaks B and C as well as the width 
of peak C.\@ Figure \ref{fig:Quanty_energy_levels} shows the low-energy excitation energies of 
a Ta$_4$ tetrahedron, in particular for intra-$t_2$ and $e$-to-$t_2$ excitations. 
The first panel on the left depicts the effect of hopping $t_\sigma$ and $t_\pi$ 
for a fixed ratio $|t_\sigma/t_\pi|$\,$\approx$\,2.6, i.e., $\alpha$\,=\,2.9, 
while the second panel introduces spin-orbit coupling $\zeta$. 
Both panels employ H$_{e-e}$\,=\,0, i.e., the non-interacting scenario. 
For $\zeta$\,=\,0, the ground state configuration is within the $a_1^2 e^4 t_2^1$ 
manifold, while the lowest excited states correspond to $a_1^2 e^3 t_2^2$. 
Finite $\zeta$ yields a $J_{\rm tet}$\,=\,3/2 ground state and a $J_{\rm tet}$\,=\,1/2 
doublet at an energy of roughly $\zeta$, i.e., the excitation energy is reduced from 
the expectation 1.5\,$\zeta$ for $\alpha$\,=\,$\infty$, see Fig.\ 3b) of the main text. 
Figure 3b) also shows that the renormalization is slightly different compared to our 
single-particle calculations and that the normalization depends on the set of orbitals 
to which spin-orbit coupling is restricted. 
At higher energy, $\zeta$ splits the $a_1^2 e^3 t_2^2$ manifold into three energy levels 
that, in the absence of electron-electron interactions, can be distinguished according to 
the occupation of the $t_2$ states: $J_{\rm 3/2}^2$, $J_{\rm 3/2}^1\,J_{\rm 1/2}^1$, 
and $J_{\rm 1/2}^2$. 
Note that the latter state cannot be reached from a $J_{\rm 3/2}^1$ ground state 
in a single-particle scenario. The two lower branches hence correspond to peaks B and C.\@ 
Electron-electron interactions cause a fanning out of energies for the $a_1^2 e^3 t_2^2$ states 
but have little effect on the low-energy $a_1^2 e^4 t_2^1$ states, i.e., $J_{\rm tet}$\,=\,3/2 
and 1/2. Based on the RIXS peak energies and the $\mathbf{q}$ dependence, we employ 
$\alpha$\,=\,2.9, $t_\pi$\,=\,0.45\,eV, $t_\sigma$\,$\approx$\,1.17\,eV, 
$\zeta$\,=\,0.27\,eV, $U$\,=\,1.75\,eV, and $J_H$\,=\,0.4\,eV.\@ 
These parameters have also been used for Fig.\ 3b) of the main text, 
where $\alpha$ has been changed by varying $t_\sigma$.

The single-particle model offers a very good description of peak A, 
the excitation from $J_{\rm tet}$\,=\,3/2 to 1/2 within the $a_1^2 e^4 t_2^1$ manifold. 
This result is stable against the addition of electron-electron interactions. 
For $\alpha$\,=\,2.9, Fig.\ \ref{fig:Quanty_spectra_map_intergral} compares the $\mathbf{q}$-dependent 
intensity of peak A along (7.35\,\,7.35\,\,$l$) for vanishing, intermediate, and sizable 
electron-electron interactions. For all three curves, spin-orbit coupling acts within 
the lower nine quasimolecular orbitals, i.e., $a_1$, $e$, $t_2$, and $t_1$. 
The behavior is very similar, and we conclude that electron-electron interactions 
have only a minor effect on peak A.

\end{document}